# Mid-Frequency Gravitational Wave Detection and Sources

Wei-Tou Ni

*National Astronomical Observatories, Chinese Academy of Sciences, Beijing, China*
*Wuhan Institute of Physics and Mathematics, Academy of Precision Measurement Science and Technology (APM), Chinese Academy of Sciences, Wuhan 430071, China*
*Department of Physics, National Tsing Hua University, Hsinchu, Taiwan, 30013, ROC*

**Abstract**
A brief history and various themes of mid-frequency gravitational wave detection are presented more or less following historical order --- Laser Interferometry, Atom Interferometry (AI), Torsion Bar Antenna (TOBA), and Superconducting Omni-directional Gravitational Radiation Observatory (SOGRO). Both Earth-based and Space-borne concepts are reviewed with outlook on expected astrophysical sources.

The LIGO-Virgo real-time detection of the binary black hole (BH) coalescence [1, 2, 3] and neutron star (NS) coalescence [4] opened the door of Gravitational Wave (GW) Astronomy and Multi-Messenger Astronomy. Primary BH with mass about 30 $M_\odot$ in the coalescence event were frequent observed. Together with the current LIGO/Virgo O3 Observation, more than 50 coalescence events have been discovered [5]. Specifically, NS-BH coalescence and Mass-Gap events have also been observed [5]. Parallel to the GW detection efforts of LIGO, Virgo, KAGRA [6] and their antecedents (both interferometers and resonant detectors) in the of high-frequency band (10 Hz - 100 kHz) [7], there are efforts in all other frequency bands from Hubble band (10 zHz - 10 fHz) to very high frequency band (100 kHz - 1 THz) and beyond (> 1 THz) [8]. Specifically, in the low-frequency band (100 nHz – 100 mHz), LISA [9] and ASTROD [10, 11] were proposed in 1990's among various space mission concepts. With the development of LISA and ground interferometers, the frequency sensitivity gap in the mid-frequency band (0.1 - 10 Hz) became transparent.

In the first decade of this century, many people believed that in the mid-frequency band, there are much less astrophysical GW sources since there was a theoretical mass gap, and hence, looked for detecting primordial GWs in this band. Two mission concepts – DECIGO [12] and BBO (Big Bang Observer) [13] were proposed and studied. Due to the expected low amplitude of primordial GWs at present Universe, the required sensitivity of these two mission concept is high and they belong to second-generation of GW space mission concepts or beyond.

After the first atom interferometer (AI) experiments in 1991 [14-16], there have been tremendous advances in the precision of atom interferometry – AI rotation sensors [17-19], AI gravimeters [20-25], AI gradiometers [26-29], AI measuring the Newton gravitational constant [28, 30-33] and AI measuring the fine structure constant [34-37]. Several groups have proposed to use atom interferometry for detecting GWs in the mid-frequency band and low frequency band [38-46] for both Earth-based and space borne detection.

For Earth-based GW detectors there are two common noises that have to be addressed – vibration noise and gravity gradient noise. After more than half a century of intensive study and implementation, the vibration noise can be dealt with to lower and lower frequency into the mid-frequency region. However, the gravity gradient noise (GGN) has the same effect on detectors as GW due to universality of gravitation. It has to be measured/estimated and deducted [47-50] for mid-frequency GW detection on Earth. The main limitation for GW detectors are linked to density variations from seismic activity and perturbations of the atmospheric pressure. There are 4 GW detection concepts proposed to cover the sensitivity gap in the mid-frequency band – AI, MI (Michelson Interferometer), SOGRO (Superconducting Omni-directional Gravitational Radiation Observatory) [51, 52] and TOBA (Torsion-Bar Antenna) [53-55]. Mongo concepts of AI, MI and TOBA were studied in 2013 [56]. The measurement and reduction of the GGN need 4 or 5 orders of precision to reach the sensitivity goals [56]. The Newtonian noise cancellation in full-tensor gravitational-wave detectors (including SOGRO) was studied in Ref. [57]. AI, torsion pendulums and superconducting gradiometer have been proposed to measure the gravity gradient. The AI project MIGA (Mater-wave laser Interferometric Gravitation Antenna) [58-60] has started its 200 m-prototype construction and begun to characterize Earth gravity field fluctuations for future GW detectors [60].

The scientific goals of first-generation mid-frequency GW detectors are basically to bridge the spectral gap between high-frequency and first-generation low-frequency GW sensitivities: to detect intermediate mass BH coalescence; to detect inspiral phase and predict time of stellar-mass binary black hole coalescence together with neutron star coalescence or neutron star-black hole coalescence for ground GW detectors and for multi-messenger astronomical observations; to detect compact binary inspirals for studying stellar evolution and galactic population; to detect middle frequency GWs from compact binaries falling into intermediate mass BHs for multi-band observation with ground-based GW detectors; to study the compact object population and to estimate/explore the astrophysical stochastic background from stellar object inspirals; to test general relativity and beyond-the-standard-model theories. The potential GW sources in the middle frequency band has been addressed and reviewed in [61-63]. These goals and sources are reviewed in various papers [64-68] in the special issue on Mid-frequency GW Detection and Sources of International Journal of Modern Physics D. Fig. 1 of [68] plots strain power spectral density (psd) amplitude vs. frequency for various GW detectors and GW sources.

In spite of much efforts, it is hard to observe intermediate BHs by conventional means. The observational evidence for intermediate-mass black holes (M = 100-$10^6$ M$_\odot$) is reviewed in [69]. Although there are many hints and some evidences, but no firm cases yet. After the first direct detection by LIGO [1], the situation changed. There have been frequent observations of ~30 M$_\odot$ and ~30 M$_\odot$ binary BH mergers [5]. With some mass-gap events and the frequent observation of ~30 M$_\odot$ and ~30 M$_\odot$ binary BH mergers, the mid-frequency (0.1-10 Hz) GW sources would be abundant. When the abundance of mass-gap events is determined more precisely, the sources for mid-frequency GW detection will be better estimated. *The observation and study of intermediate mass BHs is one of the major scientific goals of the mid-frequency GW detectors.*

In the special issue on Mid-frequency GW Detection and Sources of International Journal of Modern Physics D, there are 3 articles on Earth-based GW detection – SOGRO [64], TOBA [66], and ZAIGA [67], and 4 articles on space borne detection –

AIGSO [70], INO [65], AMIGO [68], and an orbit design/study of constant-arm space GW missions for mid-frequency and low-frequency GW detection [71].

As we mentioned, AI, torsion pendulum and superconducting gradiometer are three methods to measure the Earth gravity gradient fluctuations. SOGRO measures all five tensor components of the spacetime metric and enables identification of the source direction and wave polarization with a single detector, utilizes enhanced mechanical and electrical stabilities of materials at cryogenic temperatures to reject common-mode seismic noise to a very high degree, and gives an advantage in the rejection of the Newtonian noise due to its full-tensor characteristic [51, 52, 57]. In [64], all these are reviewed, discussed and updated.

The target strain sensitivity of TOBA is $10^{-19}$ Hz$^{-1/2}$ around 0.1 Hz. It assumes the use of 10 m long bars at cryogenic temperatures (4 K) with the rotations of the bars measured with Fabry–Perot cavities at the end of the bars [66]. The present Phase-III prototype under development is to complete the demonstration of noise reduction reaching a sensitivity of about $10^{-15}$ Hz$^{-1/2}$ at 0.1 Hz. Measuring the terrestrial gravity fluctuation with a sensitivity of about or below $10^{-15}$ Hz$^{-1/2}$ at 0.1 Hz is useful for two geophysical purpose – earthquake early warning and Newtonian noise reduction for GW detection. The torsion pendulum dual oscillator (TorPeDO) in Australia [72] is under development for these purposes also.

MIGA is working deep into the measuring and subtracting the gravity gradient fluctuations and an AI prototype GW detector [59, 60]. After MIGA, there are 4 AI facilities proposed – MAGIS-100 [73], ZAIGA [67], ELGAR [74] and AION [75] aiming at mid-frequency GW detection and potentially fundamental new physics. MAGIS (Matter-wave Atomic Gradiometer Interferometric Sensor) Collaboration is working toward a 100 m vertical AI prototype; their next phase is building a 1000 m device MAGIS-1000 with SURF (Sanford Underground Research Facility) lab in South Dakota as a candidate location [73]. In the third phase, MAGIS proposes to have a space mission with arm length thousands of kilometers to increase the detection sensitivity. ELGAR (European Laboratory for Gravitation and Atom-interferometric Research) [74] with an arm length goal of about 30 km in stages is aiming at their peak strain sensitivity of $4.1 \times 10^{-22}$ Hz$^{-1/2}$ at 1.7 Hz with potential choice of site at LSBB (Laboratoire Souterrain à Bas Bruit), Rustrel region, France or in Sardinia, Italy. AION (Atom Interferometer Observatory and Network) [75] proposes AION-10 (10 m arm), AION-100 (100 m arm), AION-km (km arm), and AION-space (AEDGE [76]).

Zhaoshan long-baseline Atom Interferometer Gravitation Antenna (ZAIGA) [67] is under construction and will be a new underground laser-linked atom interferometer facility. It will be in the 200-m-on-average underground of a mountain named Zhaoshan which is about 80 km southeast of Wuhan. ZAIGA will be equipped with long-baseline atom interferometers, high-precision atom clocks, and large-scale gyros. In the first phase ZAIGA facility will have 1-km horizontal tunnel (one arm) with two 1-km-apart atom interferometers as a ZAIGA-GW prototype together with a 1-km-arm-length tracking-and-ranging ZAIGA-CE-GW prototype using lattice optical clocks linked by locked lasers. In this phase, a 300-m vertical hole will also be drilled with atom fountains and atom clocks mounted for doing high-precision test of the equivalence principle of micro-particles (ZAIGA-EP) and clock-based gravitational red-shift measurement (ZAIGA-CE-R).

Radio Doppler tracking of spacecraft using precision ultra-stable clocks and microwave links [77, 78] can be considered as the first GW missions. The Doppler tracking data of the Chang'e 3 lunar mission is used recently to constrain the stochastic background of gravitational wave in cosmology within the 1 mHz to 0.05 Hz frequency band [79]. With the development of optical clocks, the precision has increased. Dedicated GW mission concepts using precision optical clocks and laser links have been proposed [80-82, 65]. INO [65] proposes to locate three spacecraft at one AU distance (say at L1, L4 and L5 of the Sun-Earth orbit), and to apply the Doppler tracking method for obtaining the sensitivity for gravitational wave with third- or fourth-order improvement (in 10 μHz-1 Hz frequency band) than Cassini sensitivity.

AMIGO [68, 83] is a first-generation Astrodynamical Middle-frequency Interferometric GW Observatory. The mission concept is to use time delay interferometry for a nearly triangular LISA-like formation of 3 drag-free spacecraft with nominal arm length 10,000 km, emitting laser power 2-10 W and with telescope diameter 300-500 mm. The design GW sensitivity in the middle frequency band is $3 \times 10^{-21}$ Hz$^{-1}$. The interferometric space missions has an advantage, for the drag-free requirement (including Newtonian noise suppression) at the LISA level is demonstrated in LISA Pathfinder Mission [84, 85]. In [83], we study both geocentric and heliocentric orbit formations. For all studied solar orbit options of AMIGO, the first-generation time delay interferometry satisfies the laser frequency-noise suppression requirement. For these solar-orbit options, the acceleration to maintain the formation constant can be designed to be less than 15 nm/s$^2$ with the thruster requirement in the 15 μN range. AMIGO would be a good place to test the feasibility of the constant equal-arm option.

AIGSO proposes to use all AIs for both arm-link and phase probing and to have three spacecraft in linear formation with extension of 10 km. The three spacecraft need to maintain 5 km + 5 km constant arm-length formation [86]. In [70], the issue of orbit design and thruster requirement is studied. The acceleration to maintain the formation can be designed to be less than 30 pm/s$^2$ with the corresponding thruster requirement in the 30 nN range.

In [71], we study the orbit design, fuel and thruster requirements, and proof mass actuation requirement of B-DECIGO and DECIGO. For the geocentric orbit options of B-DECIGO which we have explored, the fuel mass requirement is a concern. For the heliocentric orbit options of B-DECIGO and DECIGO, the fuel requirement to keep the arm equal and constant would be easily satisfied. Furthermore, we explore the thruster and propellant requirements for constant arm versions of LISA and TAIJI missions and find the fuel mass requirement is not a show stopper either. The proof mass actuation noise is a concern. To have enough dynamical range, an alternate proof mass is required. Detailed laboratory study is needed and warranted.

*With the successful GW detection in high frequency band, the activities for the detection of GWs in the mid-frequency band are thriving and we can look forward to the detection.*